\documentclass[aps,prl,showpacs,twocolumn,superscriptaddress]{revtex4}
\usepackage{amsmath}
\usepackage{amssymb}
\usepackage{epsfig}
\usepackage{color}
\usepackage{graphicx,amsmath}

\begin{document}
\title{Heavy fermion spin liquid in herbertsmithite}
\author{V. R. Shaginyan}\email{vrshag@thd.pnpi.spb.ru}
\affiliation{Petersburg Nuclear Physics Institute, NRC
Kurchatov Institute, Gatchina, 188300,
Russia}\affiliation{Clark Atlanta University, Atlanta, GA
30314, USA}
\author{M. Ya. Amusia}\affiliation{Racah Institute of Physics,
The Hebrew University, 91904 Jerusalem,
Israel}\affiliation{Ioffe Physico-Technical
Institute,194021 St. Petersburg, Russia}\author{A. Z.
Msezane}\affiliation{Clark Atlanta University, Atlanta, GA
30314, USA}\author{K. G. Popov}\affiliation{Komi Science
Center, Ural Division, RAS, Syktyvkar, 167982,
Russia}\author{V. A. Stephanovich}\affiliation{Institute of
Physics, Opole University, Opole Poland}

\begin{abstract}
We analyze recent heat capacity measurements in
herbertsmithite $\rm ZnCu_3(OH)_6Cl_2$ single crystal
samples subjected to strong magnetic fields. We show that
the temperature dependence of specific heat $C_{mag}$
formed by quantum spin liquid at different magnetic fields
$B$ resembles the electronic heat capacity $C_{el}$ of the
HF metal $\rm YbRh_2Si_2$. We demonstrate that the spinon
effective mass $M^*_{mag}\propto C_{mag}/T$ exhibits a
scaling behavior like that of $C_{el}/T$. We also show that
the recent measurements of $C_{mag}$ are compatible with
those obtained on powder samples. These observations allow
us to conclude that $\rm ZnCu_3(OH)_6Cl_2$ holds a stable
strongly correlated quantum spin liquid, and a possible gap
in the spectra of spinon excitations is absent even under
the application of very high magnetic fields of 18 T.
\end{abstract}


\maketitle

Quantum spin liquids (QSLs) are promising new phases, where
exotic quantum states of matter could be realized. Although much
theoretical effort has been devoted to understand their physical
nature, the question is still far from its complete
clarification. Generally speaking, the QSL is a quantum state,
formed with hypothetic particles like fermionic spinons carrying
spin 1/2 and no charge. A number of QSLs with various types of
ground states are proposed
\cite{and,mil,herb0,herb1,herb2,herb3,herb,herb4,sl,sl1,sl2,sl3,balents:2010,sci_slg,han:2011,mend2011,han:2012,punk}
but the lack of real materials possessing them obscure the
underlying physical mechanism. On the other hand, one needs a
real theory that plays important role in the understanding and
interpreting accessible experimental facts
\cite{comm,shaginyan:2011,shaginyan:2011:C,shaginyan:2012:A}.
Measurements on magnetic insulators with geometrical frustration
produce important experimental { {data}} shedding light on the
nature of { {spinon composing QSL}}. Recent measurements
indicate that the { {insulator $\rm ZnCu_3(OH)_6Cl_2$
(herbertsmithite)}} is very likely to be the first promising
candidate to host a QSL in real bulk materials. In
herbertsmithite, the dynamic magnetic susceptibility shows that
at low temperatures quasiparticle excitations, or spinons, form
a continuum, and populate an approximately flat band crossing
the Fermi level \cite{han:2012,shaginyan:2012:A}. At the same
time, our analysis of herbertsmithite thermodynamic properties
allows us to reveal their scaling behavior. The above results
demonstrate that the properties of the herbertsmithite are
similar to those of heavy-fermion (HF) metals. Thus, it can be
viewed as a new type of strongly correlated HF electrical
insulator { {exhibiting properties of HF metals but resisting
electric current}}
\cite{shaginyan:2011,shaginyan:2011:C,shaginyan:2012:A}.

The development of this concept, however, faces fundamental
problems still remaining to be resolved. The first one is that
the experimental data are taken in measurements on
herbertsmithite powder samples
\cite{herb1,herb2,herb3,shaginyan:2011,shaginyan:2011:C,shaginyan:2012:A}.
As a result, both out-of-plane magnetic defects and site
disorder between the $\rm Cu$ and $\rm Zn$ ions can strongly
change the behavior of the bulk susceptibility or even alter its
{ {low-temperature scaling at weak magnetic fields, see e.g.
Ref. \cite{mend2011}.}} The second problem is based on
experimental data showing that QSL in the herbertsmithite is
unstable against the application of external magnetic fields.
This instability is represented by { {the emergence of so-called
spin-solid phase, which, in turn, is separated from the
spin-liquid one by a gap induced by an applied magnetic field
\cite{jeong:2011}. Thus, the experimental observations of QSL
remain scattered and need regimentation. In our opinion, this
can be partially achieved by presenting a reliable
interpretation of the recent measurements of specific heat $C$
in high magnetic fields performed on herbertsmithite single
crystal samples \cite{t_han:2012,t_han:2014}.}}

In this Communication we analyze the recent heat capacity $C$
measurements in strong magnetic fields on herbertsmithite. { {We
interpret the above measurements in terms of QSL spinon
contribution. To obtain the spinon - determined partial specific
heat $C_{mag}$, we subtract the phonon and Schottky effect
contributions from the total heat capacity $C$. It turns out
that obtained $C_{mag}$ as a function of temperature $T$ at
fixed magnetic field $B$ behaves very}} similarly to the
electronic specific heat $C_{el}$ of the HF metal $\rm
YbRh_2Si_2$. This observation allows us to conclude that spinons
form the Fermi sphere with the Fermi momentum $p_F$, while the
gap in the spectra of spinon excitations and the spin-solid
phase are absent. We demonstrate that the effective mass of
spinon $M^*_{mag}\propto C_{mag}/T$ exhibits a scaling behavior
resembling that of $C_{el}/T$. We also show that new
measurements of $C_{mag}$ on single crystal samples are
compatible with those obtained on powder samples.

\begin{figure} [! ht]
\begin{center}
\includegraphics [width=0.47\textwidth]{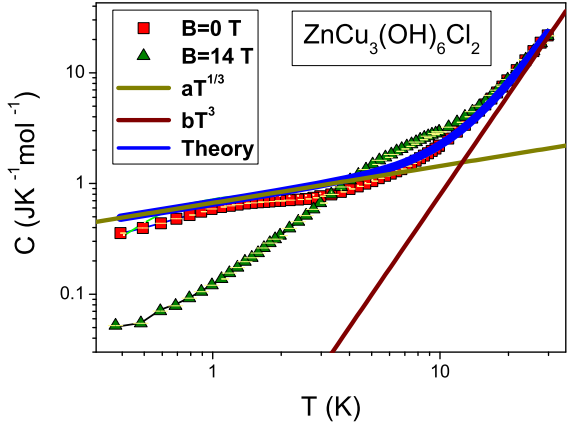}
\end{center}
\vspace*{-1.0cm} \caption{(color online). {The heat
capacity measurements on single crystal of $\rm
ZnCu_3(OH)_6Cl_2$ at $B=0$ (squares) and $B=14$ T
(triangles) \cite{t_han:2012,t_han:2014}. The thick blue
line corresponds to our theoretical approximation
\eqref{CT} with fitting parameters $a$ and $b$. The lines
represent $C=aT^{1/3}$ and $C=bT^3$, see legend.}
}\label{fig1p}
\end{figure}
\begin{figure} [! ht]
\begin{center}
\includegraphics [width=0.47\textwidth]{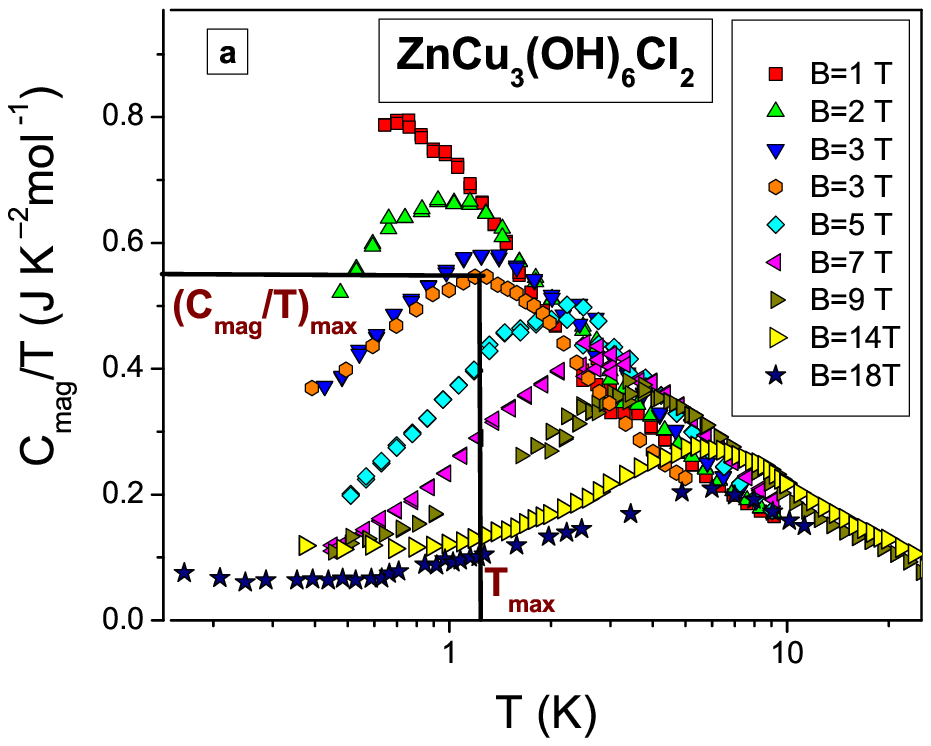}
\includegraphics [width=0.47\textwidth]{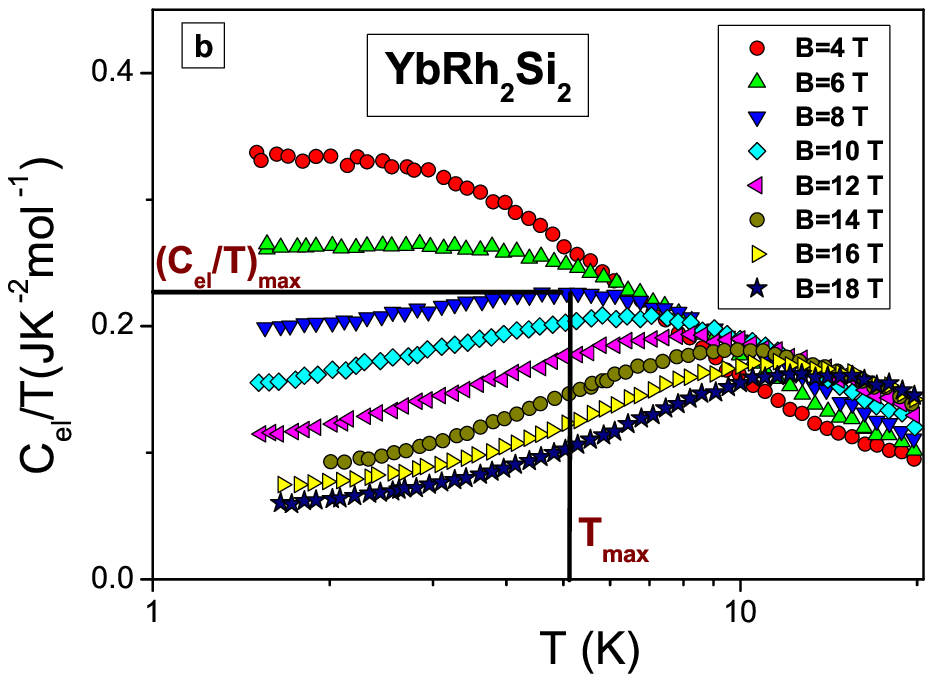}
\end{center}
\vspace*{-0.7cm} \caption{(color online). Panel a: The
specific heat { {$C_{mag}/T$ of $\rm ZnCu_3(OH)_6Cl_2$
measured on powder \cite{herb2} and single crystal samples
\cite{t_han:2012,t_han:2014}. $(C_{mag}/T)$ is displayed as
a function of temperature $T$ for fields $B$ shown in the
legends. }} Panel b: The specific heat $C_{el}/T$ extracted
from measurements on the HF metal $\rm YbRh_2Si_2$
\cite{gegenwart:2006:A} exhibits the same behavior as
$C_{mag}/T$. The illustrative values of the maxima
$(C_{mag}/T)_{\rm max}$ and $(C_{el}/T)_{\rm max}$ of both
$C_{mag}/T$ and $C_{el}/T$ and the corresponding $T_{\rm
max}$ are also shown in panels a and b.}\label{fig1}
\end{figure}

The $S=1/2$ spins of the $\rm Cu^{+2}$ ions occupy { {the
positions in a highly}} symmetric kagome lattice. As spins
on the { {above}} lattice reside in a highly symmetric
structure of corner-sharing triangles, the ground state
energy does not depend on the spins configuration. As a
result, spins located at the kagome hexagon, composed of
the two triangles, form a (frustrated) pattern that is even
more frustrated than the triangular lattice considered by
Anderson \cite{lee:2008}. As a result, the kagome lattice
has a flat topologically protected branch of the spectrum
with zero excitation energy
\cite{green:2010,heikkila:2011}. Therefore, the fermion
condensation quantum phase transition (FCQPT) can be
considered as a quantum critical point (QCP) of the $\rm
ZnCu_3(OH)_6Cl_2$ { {QSL}} \cite{shaginyan:2011,pr}. In
that case, we expect that at low temperatures the heat
capacity $C$ of herbertsmithite becomes a function of the
applied magnetic field, as it is seen from Fig.
\ref{fig1p}. We propose that { {mentioned QSL}} is composed
of chargeless fermions called spinons with the effective
mass $M^*_{mag}$. { {Latter fermions with}} spin
$\sigma=1/2$ occupy the corresponding Fermi sphere with the
Fermi momentum $p_F$, and form the excitation spectrum
typical for HF liquid located near FCQPT, while spinons
represent HF quasiparticles of deconfined QSL. Thus, the
ground state energy $E(n)$ is given by the Landau
functional depending on the spinon distribution function
$n_\sigma({\bf p})$, where ${\bf p}$ is the momentum. Near
FCQPT point, the effective mass $M^*_{mag}$ is governed by
the Landau equation \cite{land,pr,book}
\begin{eqnarray}
\label{HC3}
&&\frac{1}{M^*_{mag}(T,B)}=\frac{1}{M^*_{mag}(T=0,B=0)}\\&+&
\frac{1}{p_F^2}\sum_{\sigma_1}\int\frac{{\bf p}_F{\bf p_1}}{p_F}
F_{\sigma,\sigma_1}({\bf p_F},{\bf p}_1)\frac{\partial\delta
n_{\sigma_1}({\bf p}_1)} {\partial{p}_1}\frac{d{\bf p}_1}{(2\pi)
^3}.\nonumber
\end{eqnarray}
Here we have rewritten the spinon distribution function as
$\delta n_{\sigma}({\bf p})\equiv n_{\sigma}({\bf
p},T,B)-n_{\sigma}({\bf p},T=0,B=0)$. The Landau
interaction $F$ is defined by the fact that the system has
to be at FCQPT. Thus, { {this interaction}} brings the
system to FCQPT point, where Fermi surface alters its
topology so that the effective mass acquires temperature
and field dependence \cite{pr,ckz,khodb}. At this point,
the term $1/M^*_{mag}(T=0,B=0)$ vanishes and Eq.
\eqref{HC3} becomes homogeneous. It can then be solved
analytically. At $B=0$, the effective mass depends on $T$
demonstrating the NFL (non-Fermi liquid) behavior
\begin{equation}
M^*_{mag}(T)\simeq a_TT^{-2/3}.\label{MTT}
\end{equation}
At finite $T$, the application of magnetic field $B$ drives
the system to Landau Fermi liquid (LFL) region with
\begin{equation}
M^*_{mag}(B)\simeq a_BB^{-2/3}.\label{MBB}
\end{equation}
At finite $B$ and $T$ near FCQPT, the solutions of Eq.
\eqref{HC3} $M^*_{mag}(B,T)$ can be well approximated by a
simple universal interpolating function. The interpolation
occurs between the LFL ($M^*(T)\propto const)$ and NFL
($M^*_{mag}(T)\propto T^{-2/3}$) regions. As it is seen from Eq.
\eqref{MBB}, under the application of magnetic field $M^*_{mag}$
becomes finite, and at low temperatures the system demonstrates
the LFL behavior $C_{mag}(T)/T\propto M^*_{mag}(T)\simeq
M^*_{mag}(T=0)+a_1T^2$. Then, as it is seen from Fig.
\ref{fig1}, at increasing temperatures $M^*_{mag}\propto
C_{mag}/T$ grows, and enters a transition regime, reaching its
maximum $M^*_{\rm max}\propto (C_{mag}(T)/T)_{\rm max}$ at
$T=T_{\rm max}$, with subsequent diminishing given by Eq.
\eqref{MTT}.

To reveal a scaling behavior, we introduce the normalized
effective mass $M^*_N$ and the normalized temperature $T_N$
dividing the effective mass $M^*_{mag}$ by its maximal
values, $M^*_{\rm max}$, and temperature $T$ by $T_{\rm
max}$ at which the maximum occurs. The normalized effective
mass $M^*_N=M^*/M^*_{\rm max}$ as a function of the
normalized temperature $y=T_N=T/T_{\rm max}$ is given by
the interpolating function that approximates the solution
of Eq. \eqref{HC3}\, \cite{pr,book}
\begin{equation}M^*_N(y)\approx c_0\frac{1+c_1y^2}{1+c_2y^{8/3}}.
\label{UN2}
\end{equation}
Here $c_0=(1+c_2)/(1+c_1)$, $c_1$ and $c_2$ are fitting
parameters, $c_1$ and $c_2$ are fitting parameters,
parameterizing the Landau interaction. Magnetic field $B$
enters Eq. \eqref{HC3} only in the combination
$\mu_BB/k_BT$, making $k_BT_{\rm max}\simeq \mu_BB$ where
$k_B$ is Boltzmann constant and $\mu_B$ is Bohr magneton
\cite{ckz,pr}. Thus, Eq. \eqref{UN2} exhibits the scaling
behavior, { {which is the}} intrinsic property of Eq.
\eqref{HC3} at FCQPT: $M^*_N$ at different magnetic fields
merges into a single { {curve}} as a function of the
variable $y$. In what follows we use Eq. \eqref{UN2} to
clarify our calculations based on Eq. \eqref{HC3}.

\begin{figure}[!ht]
\vspace*{-0.3cm}
\begin{center}
\includegraphics [width=0.47\textwidth]{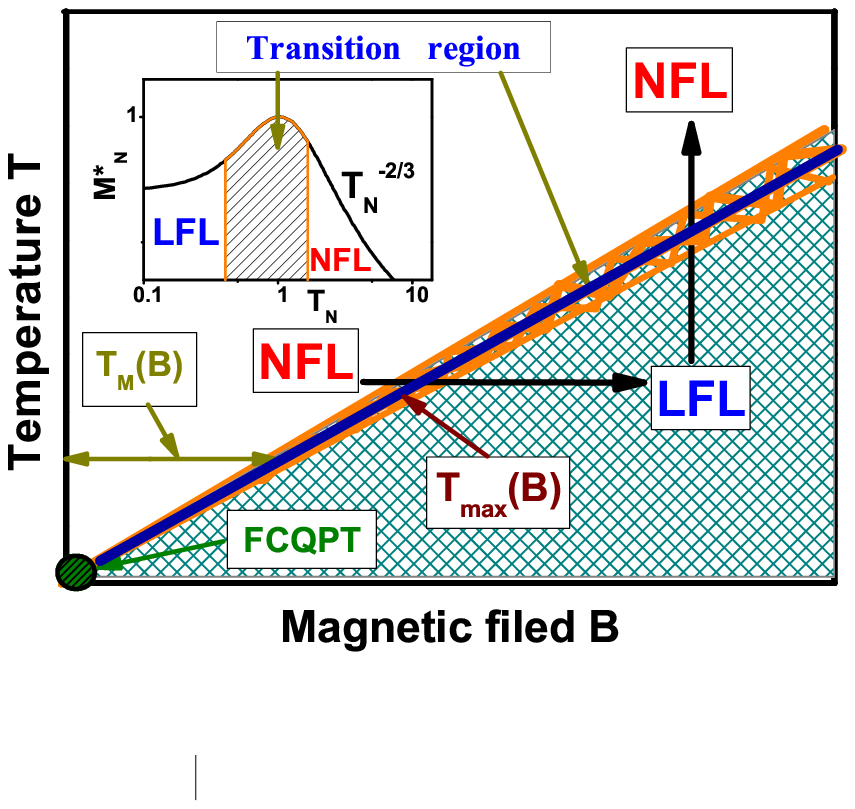}
\vspace*{-0.5cm}
\end{center}
\vspace*{-1.5cm} \caption{(color online). Schematic $T-B$
phase diagram of SCQSL with magnetic field as control
parameter. The vertical and horizontal arrows show LFL-NFL
and reverse transitions at fixed $B$ and $T$, respectively.
The { {hatched areas in the inset and main panel}} shown by
the arrows represent the transition region at $T_{\rm
max}(B)$. The solid line in the hatched area represents the
function $T_{\rm max}(B)\sim T_M(B)$. The functions
$T_M(B)$ shown by two-headed arrows define the total widths
of the NFL state and the transition area. The inset shows a
plot of the normalized effective mass $M^*_N$ versus the
normalized temperature $T_N$. Transition region, where
$M^*$ reaches its maximum $M^*_{\rm max}$ at $T_N=T/T_{\rm
max}=1$, is shown by the arrow.}\label{fig3}
\end{figure}
To analyze the { {spinon}} specific heat $C_{mag}(T,B)$, we
have to exclude the contribution coming form lattice
(phonons) that is independent of $B$. To separate $C_{mag}$
contribution, we approximate $C(T,B=0)$ by the function
\begin{equation}\label{CT}
C(T)=aT^{1/3}+bT^3.
\end{equation}
Here the first term proportional { {to $M^*_{mag}(T)\propto
T^{1/3}$}} presents the QSL contribution when it exhibits the
NFL behavior, as it follows from Eq. \eqref{MTT}, while the
second one is determined by the { {lattice (phonon)}}
contribution. It is seen from Fig. \ref{fig1p}, that the
approximation \eqref{CT} is valid in a wide temperature range.
At low temperatures $T\leq4$ K the heat capacity is mainly
formed by spinons, while the phonon contribution prevails at
higher temperatures, making its main contribution at $T\geq10$
K. It is seen, that at diminishing temperatures the LFL
behavior, $C\propto T$, starts to form. We note, that the
Schottky anomaly was removed, as it is usually done, see e.g.
\cite{t_han:2012,t_han:2014}. Upon comparing the data on the
heat capacity taken at $B=9$ and $B=14$ T shown in Fig.
\ref{fig1p}, one concludes that $C$ strongly depends on the
magnetic field. However, at raising temperatures $T\geq 10$ K
the $B$-dependence vanishes, for the contribution coming from
phonons prevails, then, the system enters the transition regime
at $k_BT\simeq \mu_BB$, and the influence of magnetic field
becomes weak, as shown in the $T-B$ phase diagram, Fig.
\ref{fig3}.

\begin{figure} [! ht]
\begin{center}
\vspace*{-0.2cm}
\includegraphics [width=0.47\textwidth]{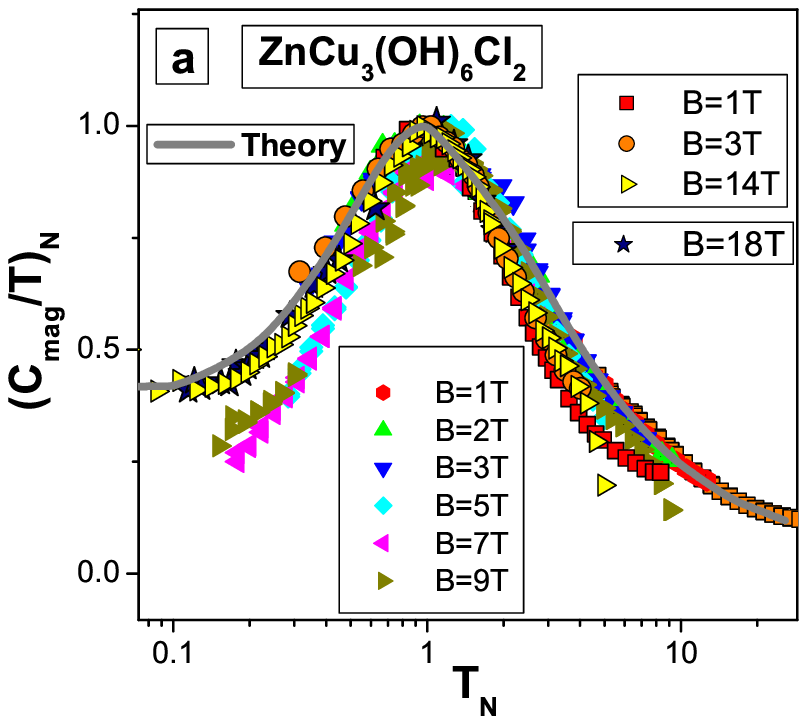}
\includegraphics [width=0.47\textwidth]{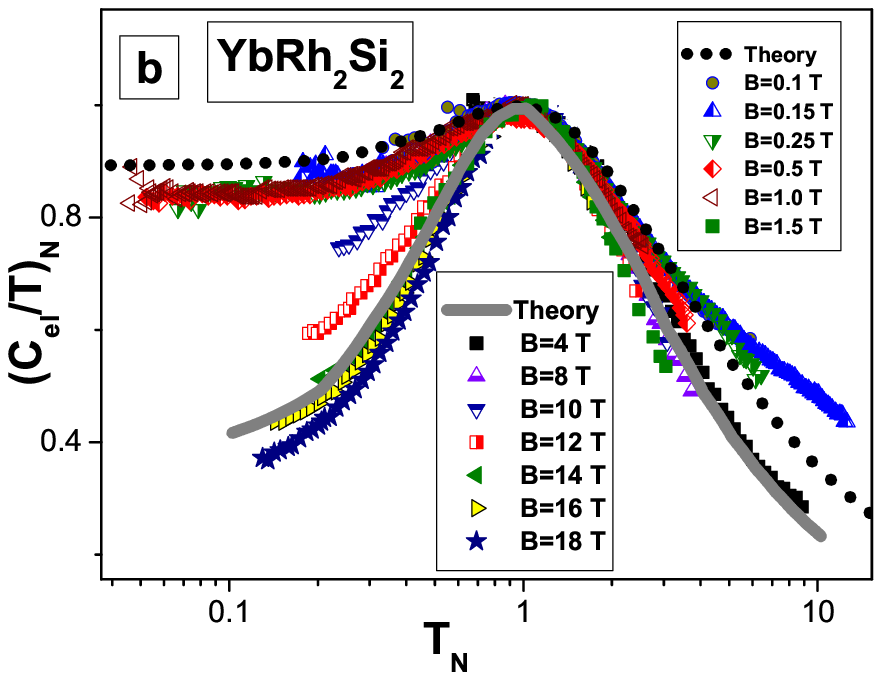}
\end{center}
\vspace*{-1.0cm} \caption{(color online). Panel a: The
normalized specific heat $(C_{mag}/T)_N=M^*_N$ of $\rm
ZnCu_3(OH)_6Cl_2$ versus normalized temperature $T_N$ as a
function of $B$ shown in the legends. $(C_{mag}/T)_N$ is
extracted from the data \cite{herb2,t_han:2012,t_han:2014}
shown in Fig. \ref{fig1}, panel a. Solid curve represents
our calculations, and traces the scaling behavior of the
effective mass. Panel b: The normalized specific heat
$(C_{el}/T)_N=M^*_N$ of $\rm YbRh_2Si_2$ versus normalized
temperature $T_N$ as a function of $B$ at low (upper box
symbols) and high (lower box symbols, corresponding { {to}}
those shown in Fig. \ref{fig1}, panel b) magnetic fields,
extracted from the specific heat measurements on the $\rm
YbRh_2Si_2$ \cite{gegenwart:2006:A,oeschler:2008}. The
low-field calculations are depicted by the dotted line
tracing the scaling behavior of $M^*_N$. The high-field
calculations (solid line) are { {performed for magnetic
fields}} when the quasiparticle band becomes fully
polarized \cite{shaginyan:2011:C1}.}\label{fig4}
\end{figure}
The obtained heat capacity of QSL, $C_{mag}/T=(C-bT^3)/T,$
is displayed in Fig. \ref{fig1}, panel a. It follows from
panel a, that the results obtained on different { {kinds
(powder and single crystal) of samples}} and in different
measurements \cite{herb1,herb2,t_han:2012,t_han:2014}
exhibit similar properties. To show that the behavior of
$C_{mag}/T$ reported in panel a is of generic character, we
note that $C_{mag}/T\propto M^*_{mag}$ behaves like
$C_{el}/T$ shown in panel b. Indeed, as seen from Fig. \ref
{fig1}, the maximum structure in $C_{mag}/T$ and $C_{el}/T$
at temperature $T_{\rm max}$ appears under the application
of magnetic field $B$ and $T_{\rm max}$ shifts to higher
$T$ as $B$ is increased. The value of the maximum structure
is saturated towards lower temperatures decreasing at
elevated magnetic fields. It is worthy to note,  that
$C_{mag}/T\sim C_{el}/T$, as it is seen from the panels a
and b. Thus, QSL of $\rm ZnCu_3(OH)_6Cl_2$ can be viewed as
a strongly correlated quantum spin liquid (SCQSL) that
behaves like that of the HF metal $\rm YbRh_2Si_2$,
exhibiting a gapless behavior in high magnetic fields up to
18 T.

\begin{figure} [! ht]
\begin{center}
\includegraphics [width=0.50\textwidth]{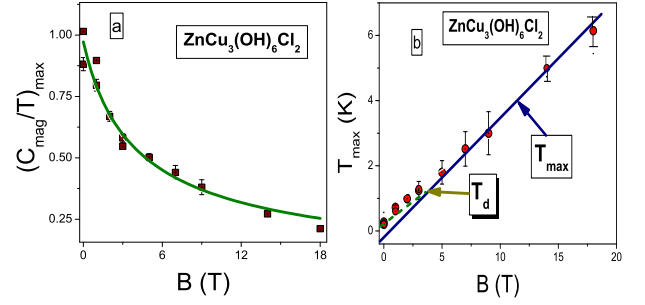}
\end{center}
\vspace*{-1.0cm} \caption{(color online). Panel a: The {
{maxima}} $(C_{mag}/T)_{\rm max}$ of $C_{mag}/T$ versus
magnetic field $B$ are shown by the solid squares, see Fig.
\ref{fig1}, panel a. The solid curve is approximated by
$M^*_{\rm max}(B)=dB^{-2/3}$, see Eq. (\ref{MBB}), where
$d_1$ is a fitting parameter. Panel b: The temperatures
$T_{\rm max}(B)$, at which the { {maxima}} of $(C_{mag}/T)$
are located, see Fig. \ref{fig1}, panel a. The solid
straight line represents the function $T_{\rm max}=d_2B$,
$d_2$ being a parameter. $T_d$ shown by the arrow is the
temperature at which the straight line $T^*_{\rm max}$
starts to deviate from the experimental data, as shown by
the dashed line.}\label{fig5}
\end{figure}
We now construct the schematic phase diagram of $\rm
ZnCu_3(OH)_6Cl_2$, { {reported}} in Fig. \ref{fig3}. At $T=0$
and $B=0$ the system is approximately located at QCP of FCQPT
without tuning. Both magnetic field $B$ and temperature $T$ play
the role of the control parameters, shifting the system from its
QCP and driving it from the NFL to LFL regions as shown by the
vertical and horizontal arrows. At fixed temperatures the
increase of $B$ leads the system along the horizontal arrow from
NFL to LFL region. This behavior is seen from Fig. \ref{fig1},
panels a and b: The application of $B$ shifts $T_{\rm max}$ to
higher temperatures. On the contrary, at fixed magnetic field
and increasing temperatures the system moves along the vertical
arrow from the LFL to NFL region. The inset to Fig. \ref{fig3}
demonstrates the behavior of the normalized effective mass
$M^*_N$ versus the normalized temperature $T_N$ that follows
from Eq. \eqref{UN2}. It is seen that the temperature { {range}}
$T_N\sim 1$ represents the transition region between the LFL {
{regime}} with almost constant effective mass and the NFL
behavior, having the $T^{-2/3}$ dependence. It follows from Eq.
\eqref{UN2} and Fig. \ref{fig3} that the width of the transition
region is given by $T_M\propto T_{\rm max}(B)\propto T\propto
B$. The phase diagram in Fig. \ref{fig3} demonstrates that the
properties of $\rm ZnCu_3(OH)_6Cl_2$ are close to those of the
HF metal $\rm YbRh_2Si_2$ \cite{pr,book}.

Figure \ref{fig4}, panel a, reports the normalized specific
heat $(C_{mag}/T)_N=M^*_N$ versus normalized temperature
$T_N$ as a function of $B$. It is seen that at low
$T_N\lesssim0.1$, $(C_{mag}/T)_N\simeq 0.4$. This value is
determined by { {the QSL polarization}} in magnetic fields,
and coincides with that of $(C_{el}/T)_N=M^*_N$ obtained on
$\rm YbRh_2Si_2$. In both panels a and b, our calculations
are shown by the same solid curve. Note that at { {low
temperatures and magnetic fields}}, when the polarization
is negligible, $(C_{el}/T)_N\simeq 0.9$, as it is seen from
panel b of Fig. \ref{fig4}.

In Fig. \ref{fig5}, panel a, the solid squares denote the
values of the maxima $(C_{mag}/T)_{\rm max}(B)$, taken from
Fig. \ref{fig1}, panel a. In panel b, the corresponding
values of $T_{\rm max}(B)$ are displayed versus magnetic
field $B$. It is seen that the agreement between the theory
and experiment is good. In panel b, the solid straight line
represents the function $T_{\rm max}(B)\propto d_2B$. It is
seen that at relatively high temperatures the experimental
data are well approximated by the straight line, while at
$T\leq T_d$ there is deviation. This fact evidences that
SCQSL of herbertsmithite is not exactly placed at FCQPT,
for in that case the approximation $T_{\rm max}\propto
d_2B$ would be good at low temperatures. This conclusion
coincides with general properties of the phase diagrams of
HF metals \cite{shag2014} and with experimental data,
showing that at low temperatures $C$ demonstrates the LFL
behavior \cite{herb2,t_han:2012,t_han:2014}, see Fig.
\ref{fig1p} as well.

In summary, based on the recent experimental facts, we have
shown that $\rm ZnCu_3(OH)_6Cl_2$ can be viewed as a
strongly correlated Fermi system whose thermodynamics is
defined by SCQSL located at FCQPT. Our calculations of the
heat capacity $C_{mag}/T$ are in good agreement with the
experimental data and the scaling behavior of $C_{mag}/T$
coincides with that observed in the HF metal $\rm
YbRh_2Si_2$ up to high magnetic fields of 18 T. Thus,
spinons forming SCQSL remain itinerant even in high
magnetic fields. We also have shown that recent
measurements of $C_{mag}$ on single crystal samples are
consistent with those obtained on powder samples.

We thank T. H. Han for valuable discussions. V.R.S. is supported
by the Russian Science Foundation, Grant No. 14-22-00281. A.Z.M.
thanks the US DOE, Division of Chemical Sciences, Office of
Energy Research, and ARO for research support. P.K.G. is partly
supported by RFBR \# 14-02-00044 and by SPbSU \# 11.38.658.2013.

\end{document}